\documentclass[conference]{IEEEtran}
\IEEEoverridecommandlockouts
\usepackage{cite}
\usepackage{amsmath,amssymb,amsfonts}
\usepackage{algorithmic}
\usepackage{graphicx}
\usepackage{textcomp}
\usepackage{xcolor}
\usepackage{subcaption}

\def\BibTeX{{\rm B\kern-.05em{\sc i\kern-.025em b}\kern-.08em
    T\kern-.1667em\lower.7ex\hbox{E}\kern-.125emX}}
\begin{document}

\title{LiDAR-based Dynamic Blockage Prediction: \\ A Data-driven Approach for Learning Interactive Bayesian Models\\
}

\author{\IEEEauthorblockN{Saleemullah Memon\textsuperscript{1,2}, Ali Krayani\textsuperscript{1}, Pamela Zontone\textsuperscript{1}, Lucio Marcenaro\textsuperscript{1}, \\ David Martin Gomez\textsuperscript{2}, and Carlo Regazzoni\textsuperscript{1}}
\IEEEauthorblockA{\textit{\textsuperscript{1}Department of Engineering and Naval Architecture (DITEN), University of Genova, Italy}\\
\textit{\textsuperscript{2}Intelligent Systems Lab, University Carlos III de Madrid, Spain}\\
saleemmemon1320@gmail.com\\
\{ali.krayani, pamela.zontone, lucio.marcenaro, carlo.regazzoni\}@unige.it, dmgomez@ing.uc3m.es}
}

\maketitle

\begin{abstract}
Vehicular sensing-based intelligence has made substantial progress in transportation systems, leading to higher levels of safety and sustainability for smart cities and
autonomous systems. This paper proposes a new approach
to learn an interactive generalized dynamic Bayesian network
(I-GDBN) model aiming to predict future LiDAR sensor
blockages from time-sequence-based 3D point cloud perception.
During learning, separate GDBN models are trained for
various vehicles in normal and blockage situations. To perform
the interaction between multiple vehicles, a high-level
vocabulary is formed. Initially, during testing, the best generative model for either normal or blockage situations is selected. An interactive Markov jump particle filter (I-MJPF) is
then proposed to leverage the probabilistic information provided by the I-GDBN 
to infer the blockages and detect the abnormalities at the high abstraction level. The proposed interactive model allows
better self-aware and explainable capabilities that can adapt to blockage scenarios, which is also helpful when sensors fail to provide observations.
\end{abstract}


\section{Introduction}
Recently, numerous research advances have been reported in the development of vehicular sensing-based intelligence relying on emerging artificial intelligence (AI), self-awareness, machine learning (ML) and deep learning (DL) approaches. The vehicles can perceive their internal states and the surrounding environment thanks to a variety of proprioceptive and exteroceptive sensors, such as GPS, cameras, RaDAR, LiDAR, and other communication devices. With the help of these sensing devices, various simple and complex tasks are investigated in the literature such as object detection \cite{R1}, localization \cite{R2, R2-b}, vehicle tracking \cite{R3}, and activity recognition \cite{R4}.
 
Sensor blockage or occlusion is a common challenge for autonomous vehicles (AVs) that has not yet been widely investigated in the literature. It mostly occurs due to situations where a sensor is obstructed or blocked by other physical objects, such as a bus, a pedestrian, or other vehicles. The situation may also occur due to sensor failure or when performing under harsh and adverse weather conditions such as wind, snow, rain, and fog. Consequently, the system is unable to properly perceive its surroundings, leading to errors or failures in navigation, continuous tracking, obstacle detection, and decision-making. An effective approach to address this challenge is by predicting those blockages before their occurrence, allowing the agent to take some proactive actions \cite{R5}. This can be made possible by the powerful capabilities of AI, ML, and DL methods which can utilize side information and prior observations to predict future blockages.

In the recent literature, some sensorial information-based blockage predictions with AI techniques have been discussed, such as a 3D lidar point cloud and mmwave communication-based convolutional neural network (CNN) model for blockage prediction is proposed in \cite{R5}. Similarly, \cite{R6} and \cite{R7} propose a vision-based CNN and recurrent neural network (RNN) combined model, and a radar sensing-based CNN and long short-term memory (LSTM) combined model, respectively, for line-of-sight blockage prediction between the base station (BS) and user. These learning approaches have high computational complexity, require large amounts of labeled data, have difficulty with real-time blockage labeling, and struggle to adapt to new unseen situations coming from the sensors. Moreover, these approaches yield a black-box nature, lack of ability to handle interactions between multiple objects, and do not possess self-awareness or explainable capabilities. The probabilistic reasoning and data-driven approaches enable adaptive learning, operate hierarchically from a low observation level to high abstraction levels, ensure explainability by modeling transitions and uncertainties from hidden patterns as well as accept multisensorial fusion, which enhances accuracy by integrating various sensor inputs \cite{R8}.

Motivated by the above discussion, we propose a novel probabilistic and data-driven approach to learning an interactive generalized dynamic Bayesian network (I-GDBN) model from time series point clouds of a 3D LiDAR sensor, where we aim to predict the LiDAR blockages and detect high-level abnormalities. For this purpose, we have performed an unsupervised growing neural gas (GNG) approach on the generalized states (GSs) and built the vocabulary for each normal GDBN model individually. For the abnormal tracks, the vocabularies are modified by considering the dummy nodes at the blockage areas. To create the global vocabulary, this paper also introduces the concept of cluster coupling and the creation of coupling variables (words) at the higher hierarchy of our proposed I-GDBN model. Moreover, during the testing phase, after selecting the best GDBN model, an interactive Markov jump particle filter (I-MJPF) is proposed to make the inference. The major contributions of this paper can be summarized as follows:

\begin{itemize}
\setlength{\itemsep}{0.2pt} 
\item To learn 3D LiDAR-based multiple GDBN models for each vehicle, detected from the joint probabilistic data association filter (JPDAF), and then learn an interaction model between them, hence to build a world model called I-GDBN.  
\setlength{\itemsep}{0.2pt} 
\item To learn the behavior of dynamic blockages by considering the dummy clusters and feeding this knowledge into our proposed I-GDBN model thus modifying its vocabulary. 
\setlength{\itemsep}{0.2pt} 
\item To choose the optimal generative model(s) for normal and blockage situations during the testing phase, which improves prediction accuracy.
\setlength{\itemsep}{0.2pt} 
\item To infer the future LiDAR sensor blockages and detect abnormalities at the interaction level from an I-MJPF. 
\end{itemize}

The remaining organization of this paper is described as follows. Section II defines the system model. Section III explains the proposed methodology. Section IV discusses the simulation results. Finally, section V draws some conclusions.

\section{System Model}
\label{sec: systemmodel}
\begin{figure}[htb]
\centering
\centerline{\includegraphics[width=8.5 cm]{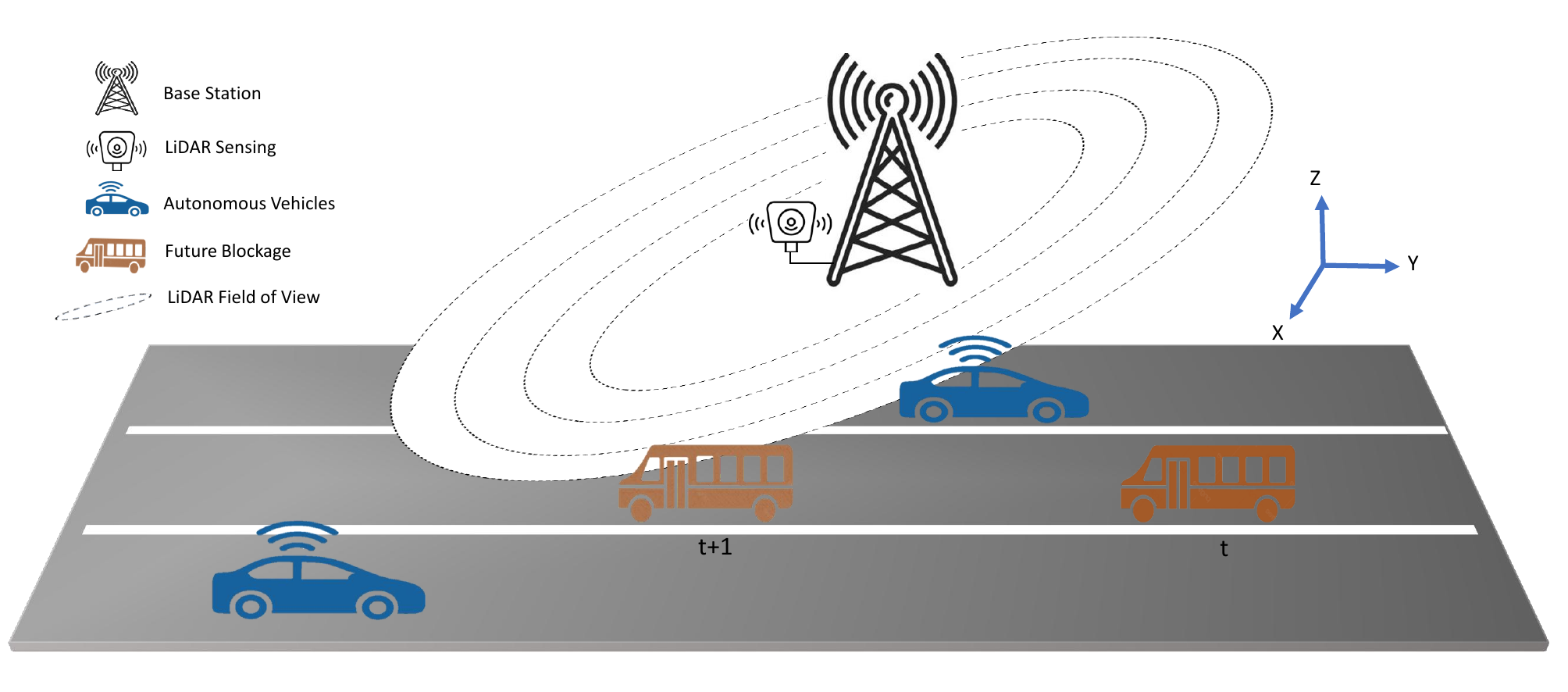}}
\caption{System model illustrating the dynamic future blockages in a dynamic environment.}
\label{fig.1}
\end{figure}

The dynamic scenario of our system model is shown in Fig.~\ref{fig.1}. It includes moving AVs which serve as wireless transmitters, and a stationary BS that acts as a wireless roadside receiver. The high-frequency BS uses a 3D LiDAR sensor which accurately perceives the environmental awareness about the static and other dynamic objects within its sensing range. The main goal of LiDAR here is to provide information about the mobility patterns, the location of the AVs and pedestrians, and the range and angle information of other objects. A bus shown in the figure, moving from right to left at time \(t\), will create a blockage very shortly at time \(t+1\), for an AV moving in the second lane from left to right side on the road. 

With the help of environmental awareness achieved by the LiDAR sensor, the BS is considered as a self-aware agent. The BS leverages this sensing information and enhances the network performance by performing continuous tracking and state prediction of the disappeared objects.
\begin{figure}[htb]
\centering
\centerline{\includegraphics[width=8.7 cm]{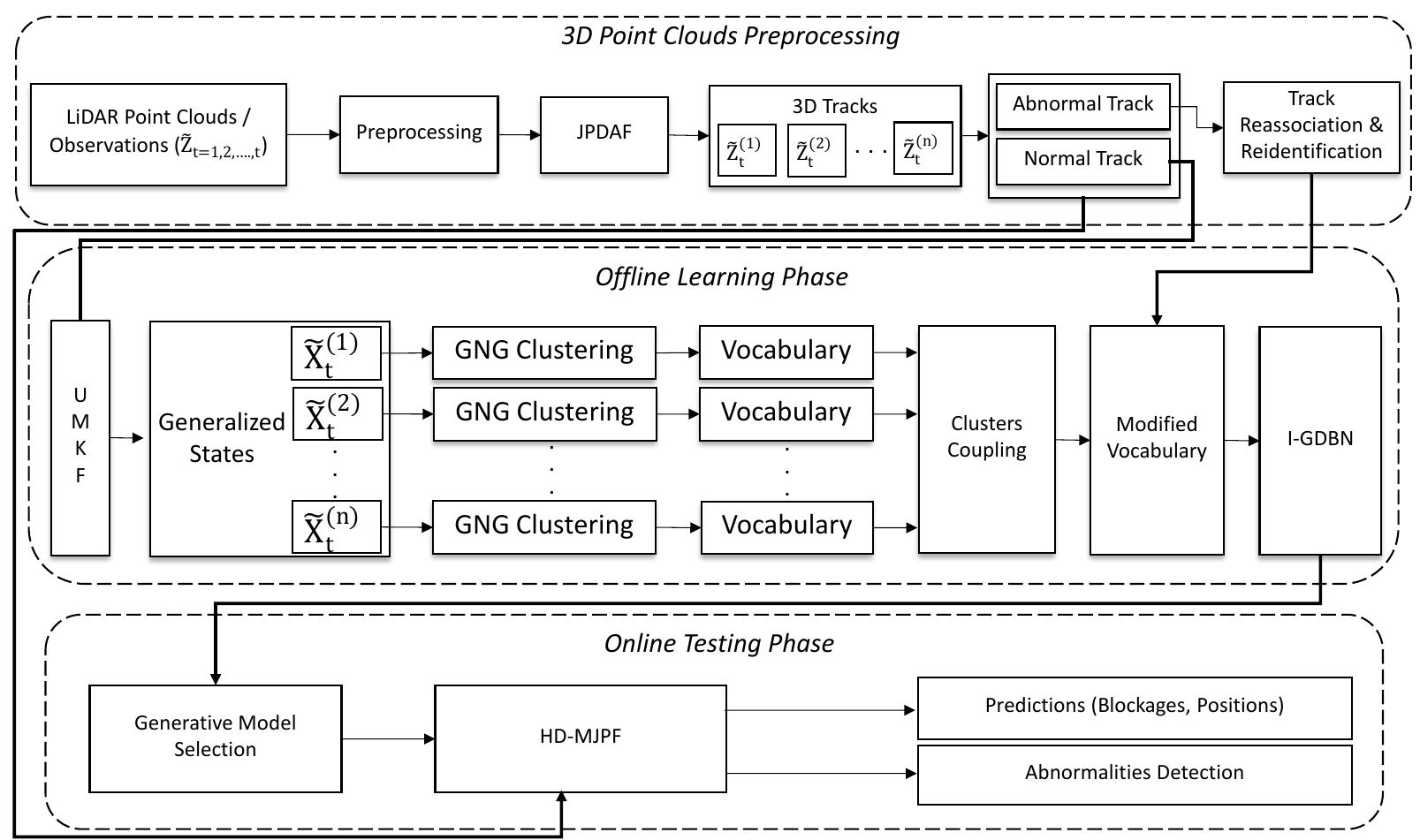}}
\caption{An illustration of the proposed methodology, showing 3D point cloud pre-processing (top),  offline training (middle), and online testing (bottom).}
\label{fig.2}
\end{figure}In this way, the agent can maintain the track's continuity, and in case of blockage or disappearance of an object, it can predict its future positions and can take certain actions before any blockage occurs. 

\section{Proposed Methodology}
\label{sec:proposedmethodology}
The proposed methodology, representing the entire process of this work, is illustrated in Fig.~\ref{fig.2}.

\subsection{Dataset Overview \& Pre-processing}
\label{ssec:preprocessing}

\begin{figure}[htb]
\begin{minipage}[b]{\linewidth}
  \centering
  \centerline{\includegraphics[width=8cm]{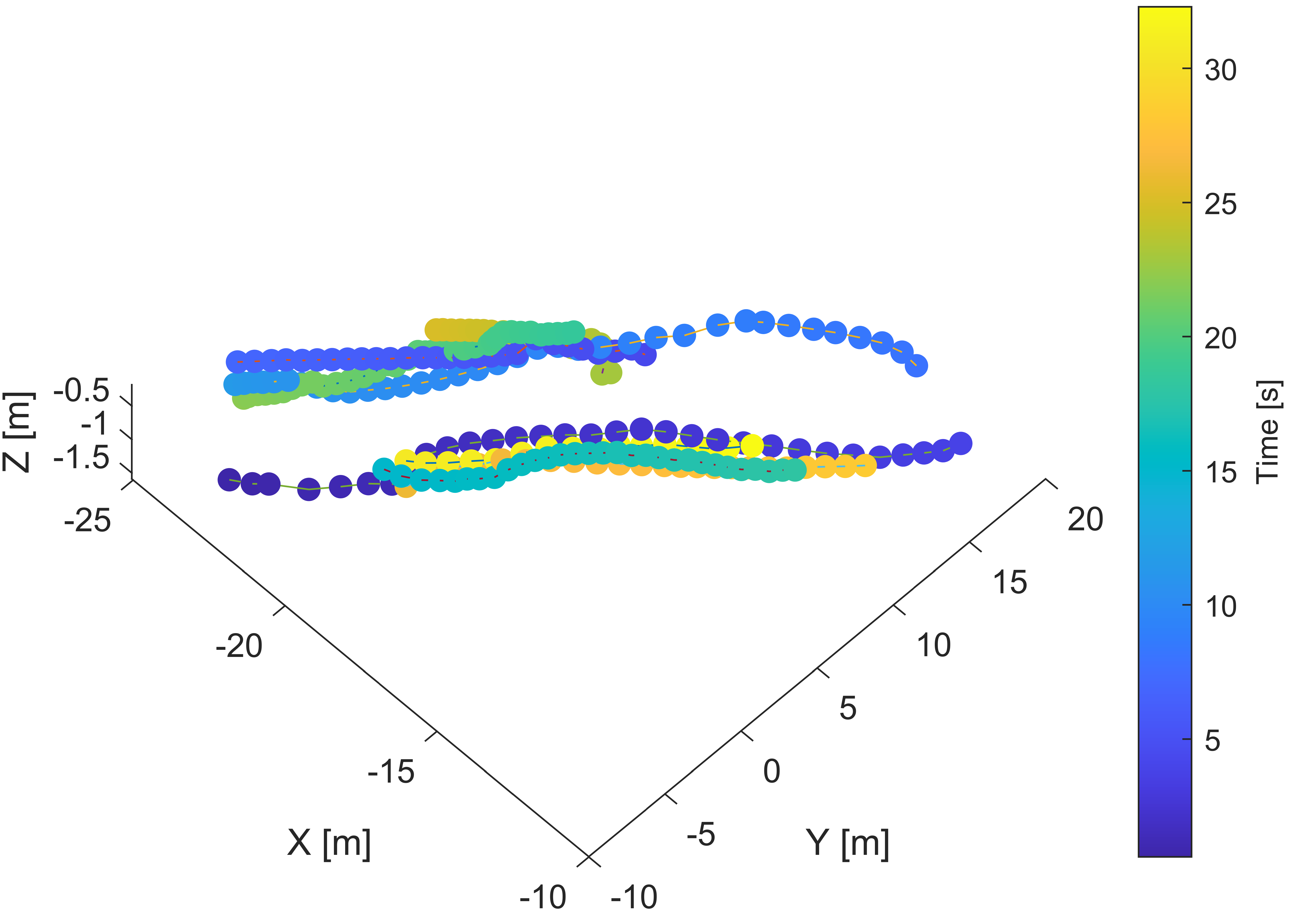}}
  \centerline{(a)}
\end{minipage}
\begin{minipage}[b]{\linewidth}
  \centering
  \centerline{\includegraphics[width=8.1cm]{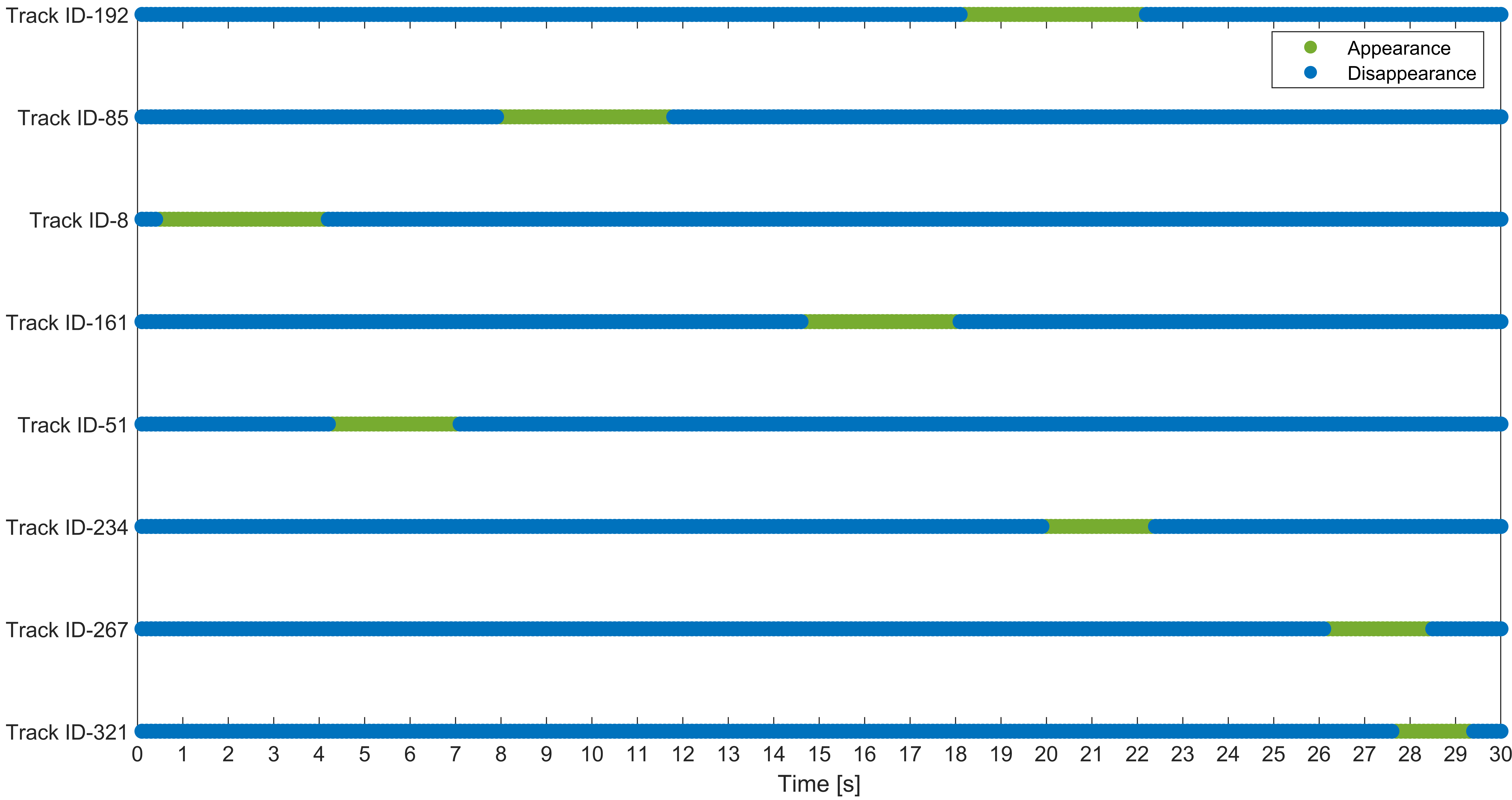}}
  \centerline{(b)}
\end{minipage}
\caption{Output of the JPDAF approach, (a) Tracking of multiple vehicles from LiDAR point clouds, (b) vehicles appearing and disappearing at different time instants.}
\label{fig.3}
\end{figure}
For both the training and testing stages, we take into consideration the deep sense 6G dataset \cite{R9}, which is based on real-world observations and is also suitable for our scenario. The dataset's street-level vehicle-to-infrastructure (V2I) communication scenario, also referred to as scenario n. 32, captures data in a wireless outdoor area, which are collected in a two-way street with two lanes at College Avenue in Arizona, USA. 
With 13-meters street width and 25-mph (40.6-kilometer-per-hour) vehicle speed limit, vehicles of various sizes move here, at varying speeds in both directions. In this work, we have only considered a time series point cloud dataset which is obtained by the exteroceptive 3D LiDAR sensor. The LiDAR has a 100-meter range, a maximum motor spin frequency of 20 Hz, and is mounted on the vehicle (which is static). 

To preprocess or extract the necessary features of dynamic objects from 3D point cloud observations \( \tilde{Z}_{t=1,2,\ldots,t} \), the raw data are first pre-processed using the LiDAR viewer toolbox in MATLAB 2024.b. The initial region of interest spans dimensions of -82 to 89 meters along the X-axis, -126 to 78 meters along the Y-axis, and -2 to 16 meters along the Z-axis. After applying filtering techniques such as ground removal, cropping, and neighbour denoising, the processed point clouds are reduced to dimensions of -28 to -4 meters along the X-axis, -13 to 35 meters along the Y-axis, and -1 to 1 meters along the Z-axis, with approximately 1000 points extracted from each point cloud.

As the pre-processed point clouds are highly dense and contain uncertainty in associating multiple observations with the corresponding dynamic objects. The JPDAF approach has been used in some applications \cite{R10}, since it better addresses the uncertainty issue, uses probabilistic weights based on prior estimates of observations, provides enhanced tracking accuracy, as well as can simultaneously detect and track various dynamic objects in the scene \cite{R3}. The JPDAF output for normal tracks is shown in Fig.~\ref{fig.3}(a). The time appearance of the detected vehicles is shown in Fig.~\ref{fig.3}(b), which is important to perform the interaction between them. The JPDAF provides two types of tracks, i.e., normal and abnormal tracks. The normal tracks are complete tracks without any blockage. The abnormal tracks instead are not continuous (they disappear from the scene due to blockage and reappear again after some time). For abnormal tracks, a reassociation and reidentification algorithm is applied to predict their missing positions. 


\subsection{Learning multiple-GDBNs}
\label{ssec:learningmGDBN}

\begin{figure}[htb]
\centering
\centerline{\includegraphics[width=8.5 cm]{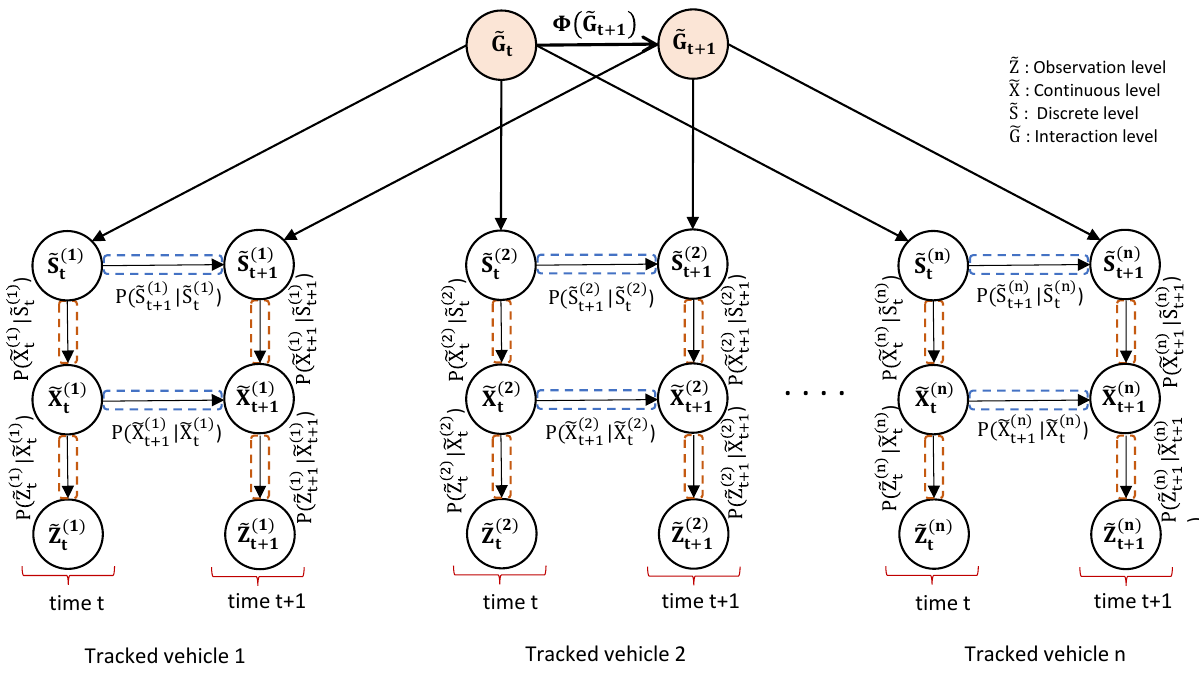}}
\caption{Learning the interaction between multiple GDBNs.}
\label{fig.4}
\end{figure}

During the learning phase of normal tracks, a separate GDBN model (i.e., the vocabulary) is trained for each vehicle, as shown in Fig.~\ref{fig.4}. An unmotivated Kalman filter (UMKF) is initially employed, which produces the GSs (\(\tilde{X}_t\)) with the filtered positions and velocities of each tracked vehicle, under the assumption that the vehicles remain unaffected by any external forces \cite{R3}. The GNG clustering is taken into consideration in order to learn the dynamic GSs of each tracked vehicle. We chose this unsupervised neural network approach because it works well in our 3D dynamic environment, does not require a fixed number of clusters, and can learn the structure of incoming data in a dynamic, incremental, and adaptable manner \cite{R11}. It yields different number of clusters for each track as an output, after receiving a 6D vector of GSs as an input. The number of clusters also depends on the size of the track. 
\begin{figure}[htb]
\begin{minipage}[b]{0.48\linewidth}
  \centering
  \centerline{\includegraphics[width=3.9cm]{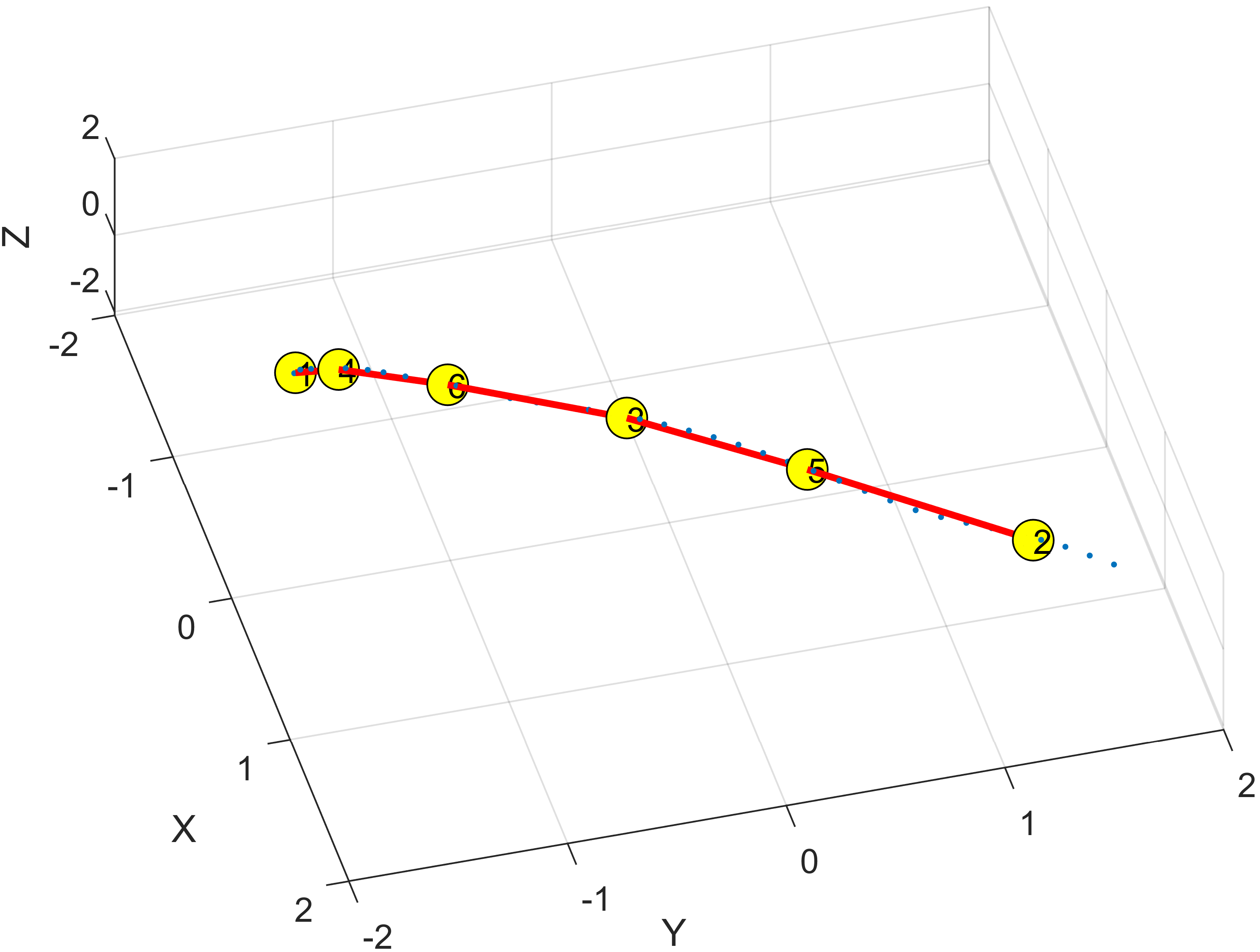}}
  \centerline{(a)}
\end{minipage}
\hfill
\vspace{0.1cm} 
\begin{minipage}[b]{0.48\linewidth}
  \centering
  \centerline{\includegraphics[width=3.8cm]{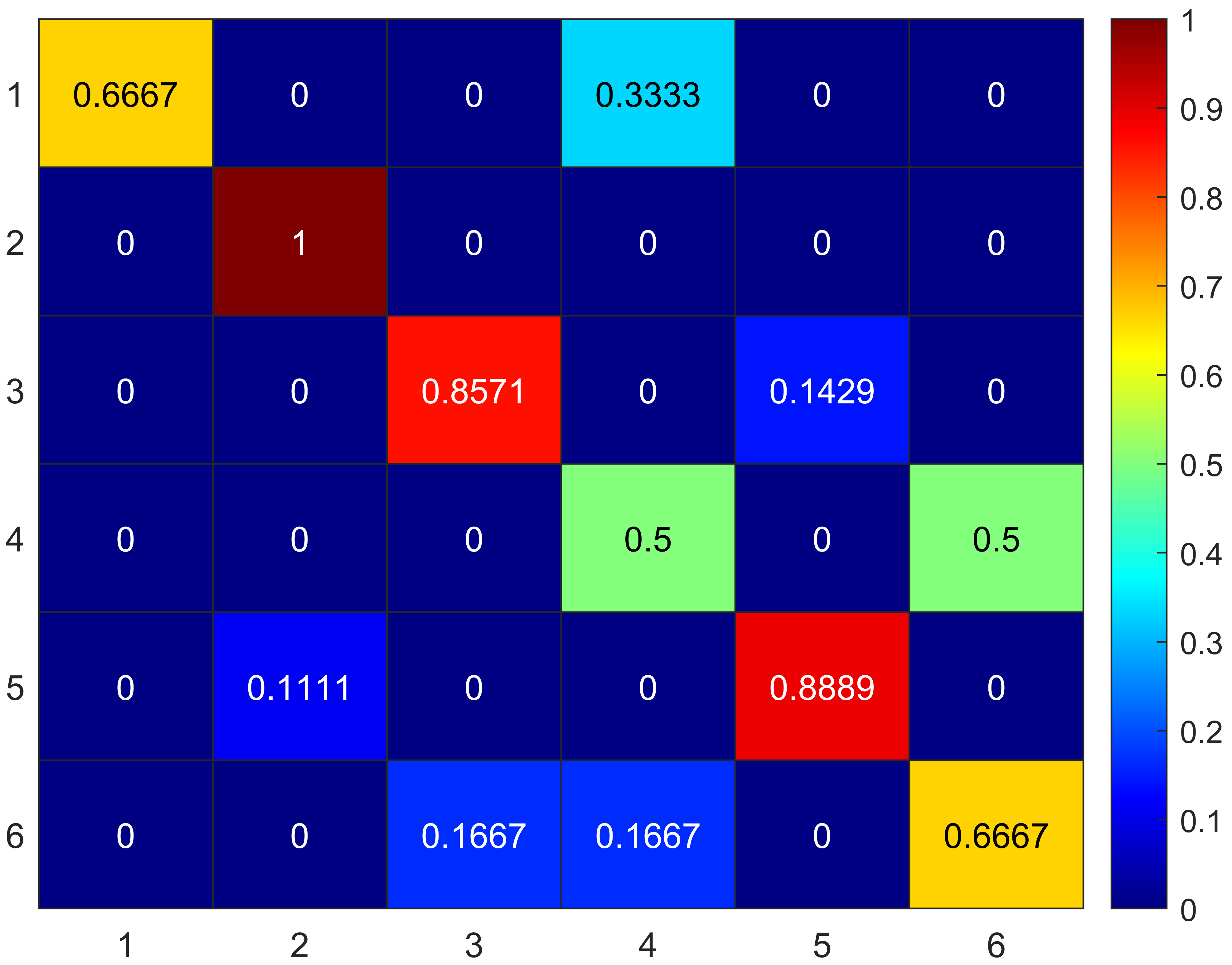}}
  \centerline{(b)}
\end{minipage}
\hfill
\begin{minipage}[b]{0.48\linewidth}
  \centering
  \centerline{\includegraphics[width=3.9cm]{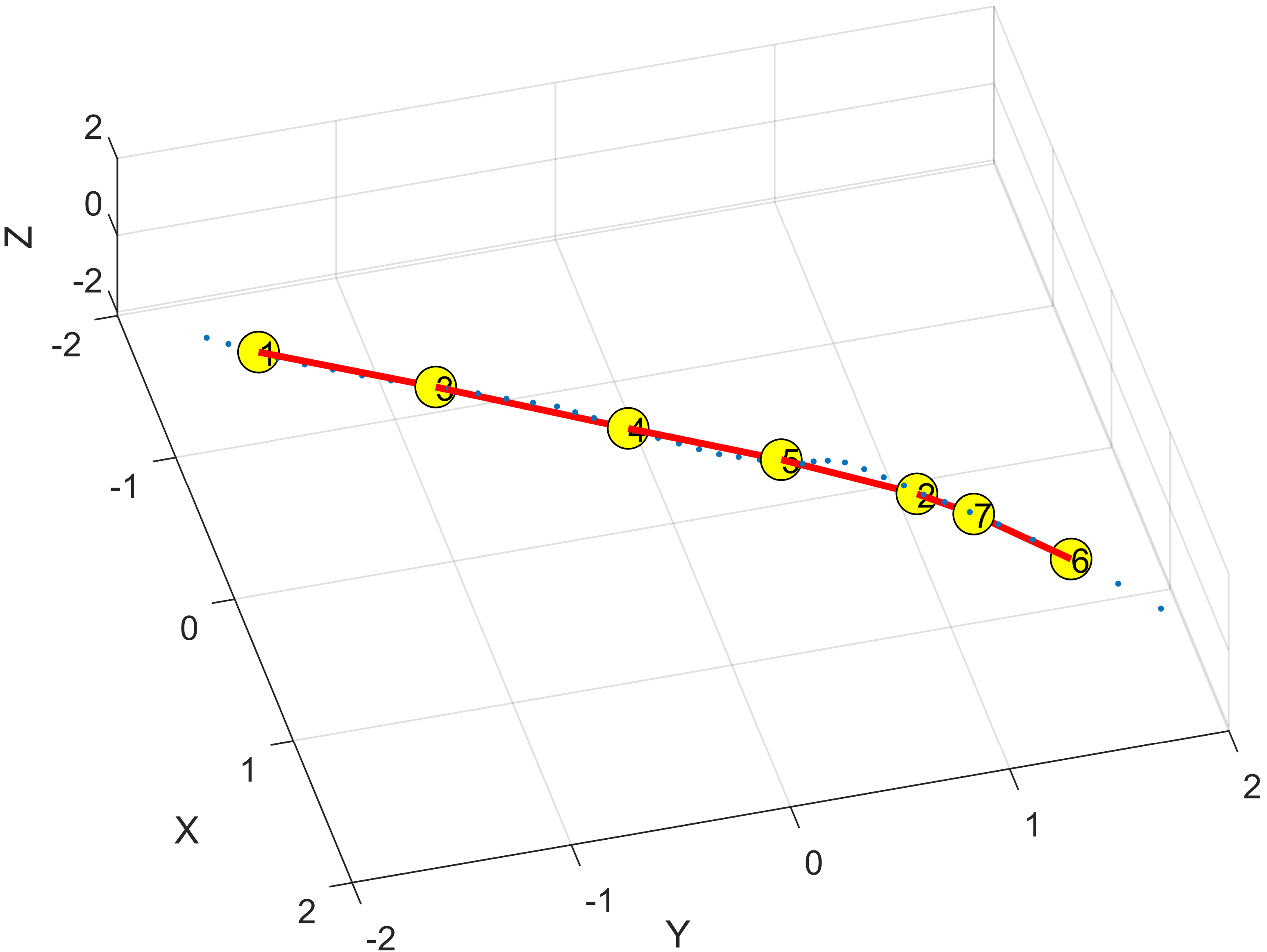}}
  \centerline{(c)}\medskip
\end{minipage}
\hfill
\begin{minipage}[b]{0.48\linewidth}
 \centering
  \centerline{\includegraphics[width=4.0cm]{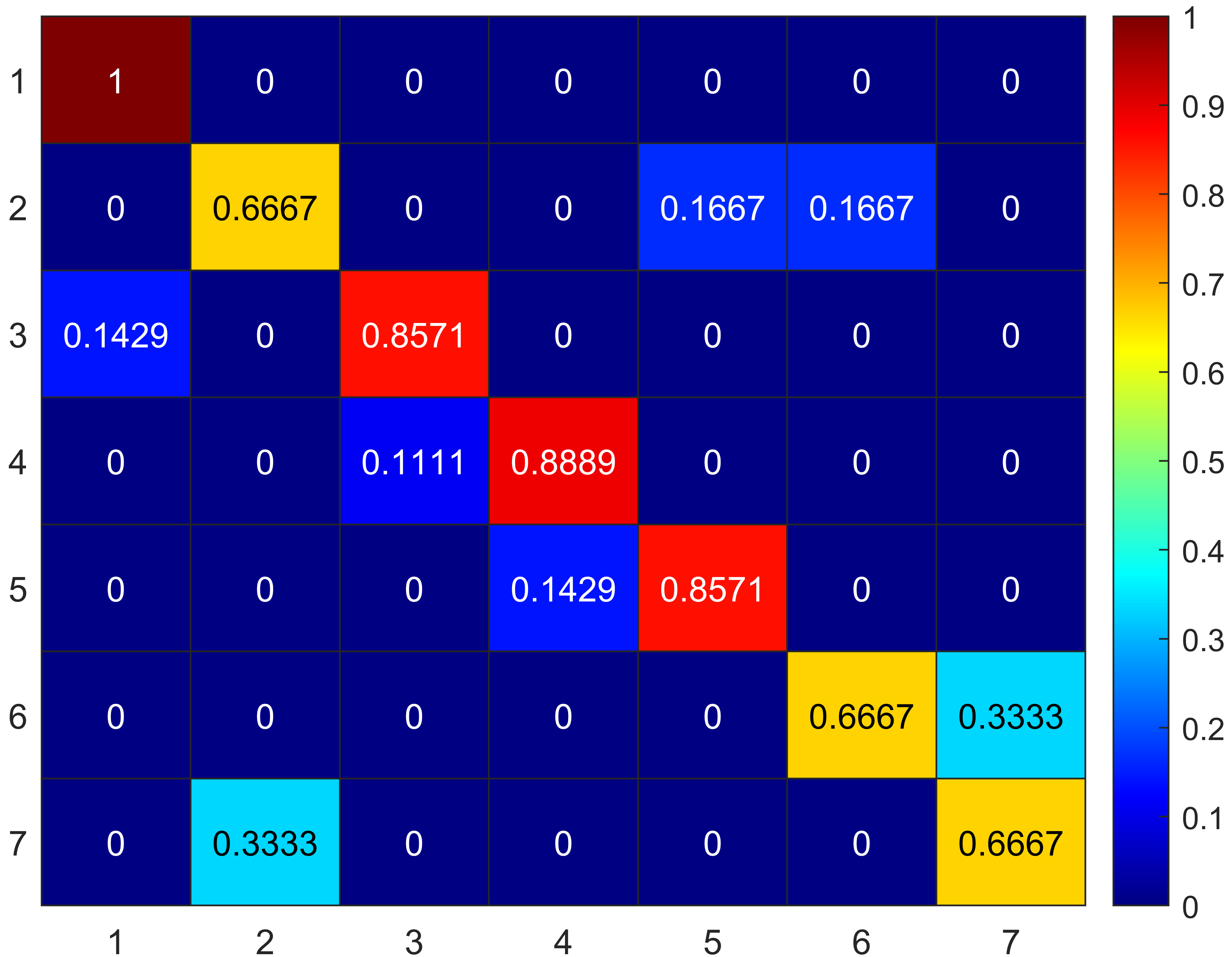}}
  \centerline{(d)}
\end{minipage}

\caption{Examples of learning from GSs (GNG clustering output): (a) Clustering for track ID-8, (b) TM for track ID-8, (a) Clustering for track ID-192, (b) TM for track ID-192.}
\label{fig.5}
\end{figure}The output of GNG, shown in Fig.~\ref{fig.5}, represents the two different examples of tracked vehicles. Each cluster represents a node or discrete variable (the discrete level (\(\tilde{S}_t\)) of GDBN). Fig.~\ref{fig.5}(a) and Fig.~\ref{fig.5}(c) show the GNG clusters of track ID-8 and track ID-192, respectively. Fig.~\ref{fig.5}(b) and Fig.~\ref{fig.5}(d) represent the transition matrices (TMs) of the corresponding track ID-8 and track ID-192, respectively. The probability of switching clusters over time is represented by these TMs. The highest probability (one) in each TM indicates the last cluster of that track, and this also provides the direction of the moving vehicle. In addition to the number of clusters and overall TM, the vocabulary of each GDBN model also contains temporal TMs, the mean (centroid), as well as the covariance matrix of each cluster. In this way, multiple vocabularies are created for multiple GDBN models. 
 For the abnormal tracks, dummy GDBN models are created with dummy clusters at the blockage positions. The TMs are also updated based on the probabilities of dummy clusters obtained from the reassociation and reidentification algorithm. In this way, the vocabularies of dummy GDBN models are modified. 

\subsection{Learning I-GDBN}
\label{ssec:learningIGDBN}

To learn the interaction among multiple GDBNs for normal and abnormal tracks, another higher abstraction hierarchy level (\(\tilde{G}_t\)) is created (see Fig.~\ref{fig.4}), and we call it an interaction or coupled level. The vehicle detection time information, obtained from the JPDAF approach, and the clusters (nodes), obtained from the GNG clustering, are taken into account for the coupling process. Here, we have defined some rules that allow us to learn a global vocabulary. Each word (\(\tilde{G}_t\)) is a combination of discrete-level clusters (\(\tilde{S}_t\)). If no vehicle is detected in the scene, then \(\tilde{G}_t=0\). The words are assigned in ascending order as time changes. For example, at a certain moment, when \(t = 4\), the two vehicles are interacting in the environment with their corresponding nodes, that are, \(\tilde{S}_4^{(1)} = 2\) and \(\tilde{S}_4^{(2)} = 3\). The assigned word, in this case, is \(\tilde{G}_4 = 10\). At the next time step, when \(t = 5\), the appearing nodes are changing as \(\tilde{S}_5^{(1)} = 3\) and \(\tilde{S}_5^{(2)} = 4\), thus the assigned word, in this case, would be \(\tilde{G}_5 = 11\). Moreover, the previous word would the remain same for the next time instant, if the vehicles remained in the same previous nodes. Besides the formation of words, the global vocabulary  also contains a global TM which represents the probabilities of moving from one word to the next word.  




\subsection{Interactive-Markov Jump Particle Filter (Online Testing)}
\label{ssec:I-MJPF}

The BS predicts the future positions of each vehicle based on a MJPF that combines the Particle filter and Kalman filter to perform temporal and hierarchical predictions. An I-MJPF is proposed here to allow more complicated interactive predictions. Before making any predictions, a suitable model will be selected for the testing track(s) by considering its 3D positions, and comparing them with the mean of each cluster for each GDBN model. For this purpose, the Euclidean distance is considered here to find the closest clusters, i.e., 
\begin{equation}
D_{kl}^{(m)} = \sqrt{(x_k - x_{c_l}^{(m)})^2 + (y_k - y_{c_l}^{(m)})^2 + (z_k - z_{c_l}^{(m)})^2}
\end{equation} 

\begin{equation}
e_k^{(m)} = \min_{l \in \{1, 2, \dots, N_m\}} D_{kl}^{(m)}
\end{equation}
\begin{equation}
E^{(m)} = \sum_{k=1}^{M} e_k^{(m)}
\end{equation}
\begin{equation}
m^* = \arg\min_{m \in \{1, 2, \dots, N\}}E^{(m)}
\end{equation}
where \(M\) represents the total number of 3D positions in a testing track, \(m\) indicate model number, \(N_m\) represents the total number of clusters in model \(m\), \(k\) and \(l\) denote testing positions from 1 to \(M\) and clusters mean from 1 to \(N_m\), respectively. Furthermore, \(e_k^{(m)}\) indicates the minimum errors for a track in model \(m\), and \(m^*\) symbolizes the optimum GDBN model with overall minimum errors. The closest clusters of the selected model are used to map the estimated words, thus updating the belief of our I-GDBN model.

An I-MJPF propagates a number of particles, each one representing a hypothesis regarding the current state of the system. Initially, uniform weights for each particle are assigned. To make an inference at the interaction level, the first word is randomly sampled from a probability distribution. For each particle, the next word is predicted based on the transition probabilities of its estimated word. The estimated word at the current time step is then used to update the particle weights. Particles are resampled based on their updated weights, and particles with higher weights are more likely to be selected. 

\section{Simulation Results}
\label{sec:results&discussions}
The performance of our proposed I-GDBN model is discussed in this section. The core novelty of the algorithm lies in its design, which enables the interaction among various GDBN models. We have examined the performance of our model in terms of predictions and abnormality detections at the I-GDBN’s higher level (at the word level). For the abnormality detection, we have considered the Hamming distance indicator, which is a good choice, since the predicted and estimated words are discrete. It provides the binary value ”1” when the predicted word does not match the estimated word, and the binary value ”0” when the predicted word matches the estimated word. Moreover, for better understanding of normal tracks, the same tracking IDs (obtained from the JPDAF) are assigned to the trained GDBN models. For the blockage (dummy) models, the IDs from 1000 and onwards are assigned. The prediction outcomes are represented for each time instant as the LiDAR provides 10-point cloud frames per second. \begin{figure}[htb]
\begin{minipage}[b]{0.48\linewidth}
  \centering
  \centerline{\includegraphics[width=3.65cm]{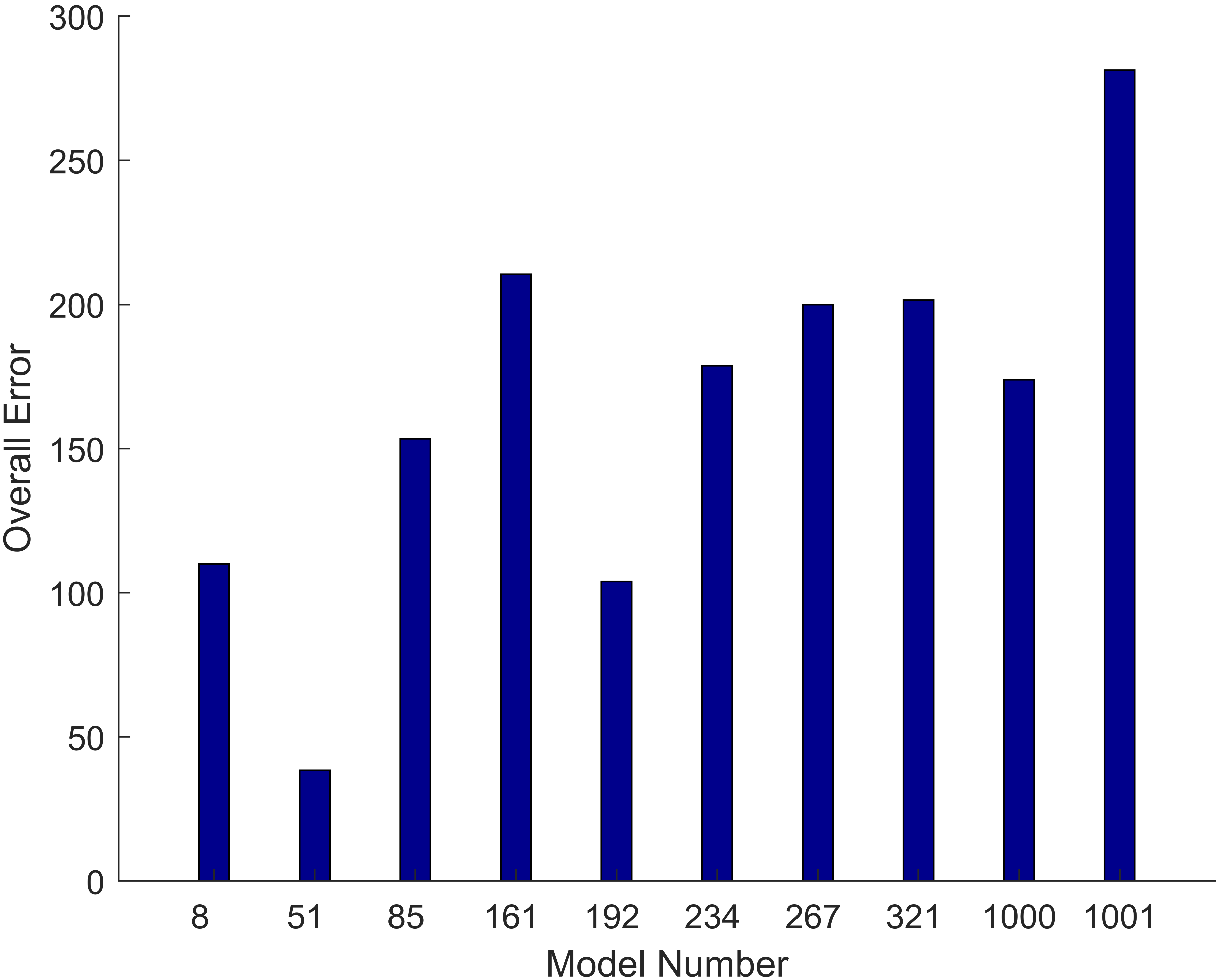}}
  \centerline{(a)}
\end{minipage}
\hfill
\begin{minipage}[b]{0.48\linewidth}
  \centering
  \centerline{\includegraphics[width=3.95cm]{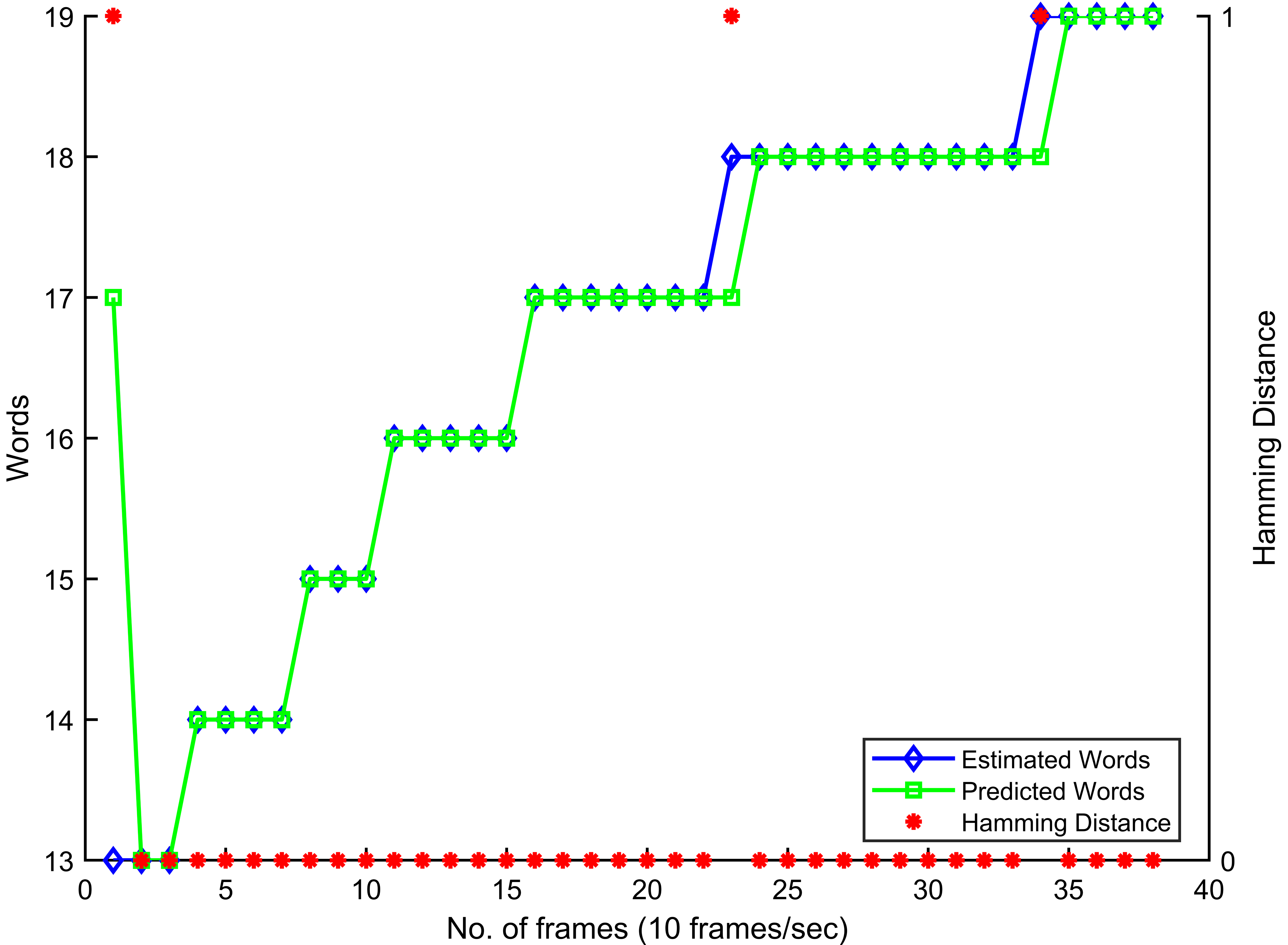}}
  \centerline{(b)}
\end{minipage}
\hfill
\begin{minipage}[b]{0.48\linewidth}
  \centering
  \centerline{\includegraphics[width=3.65cm]{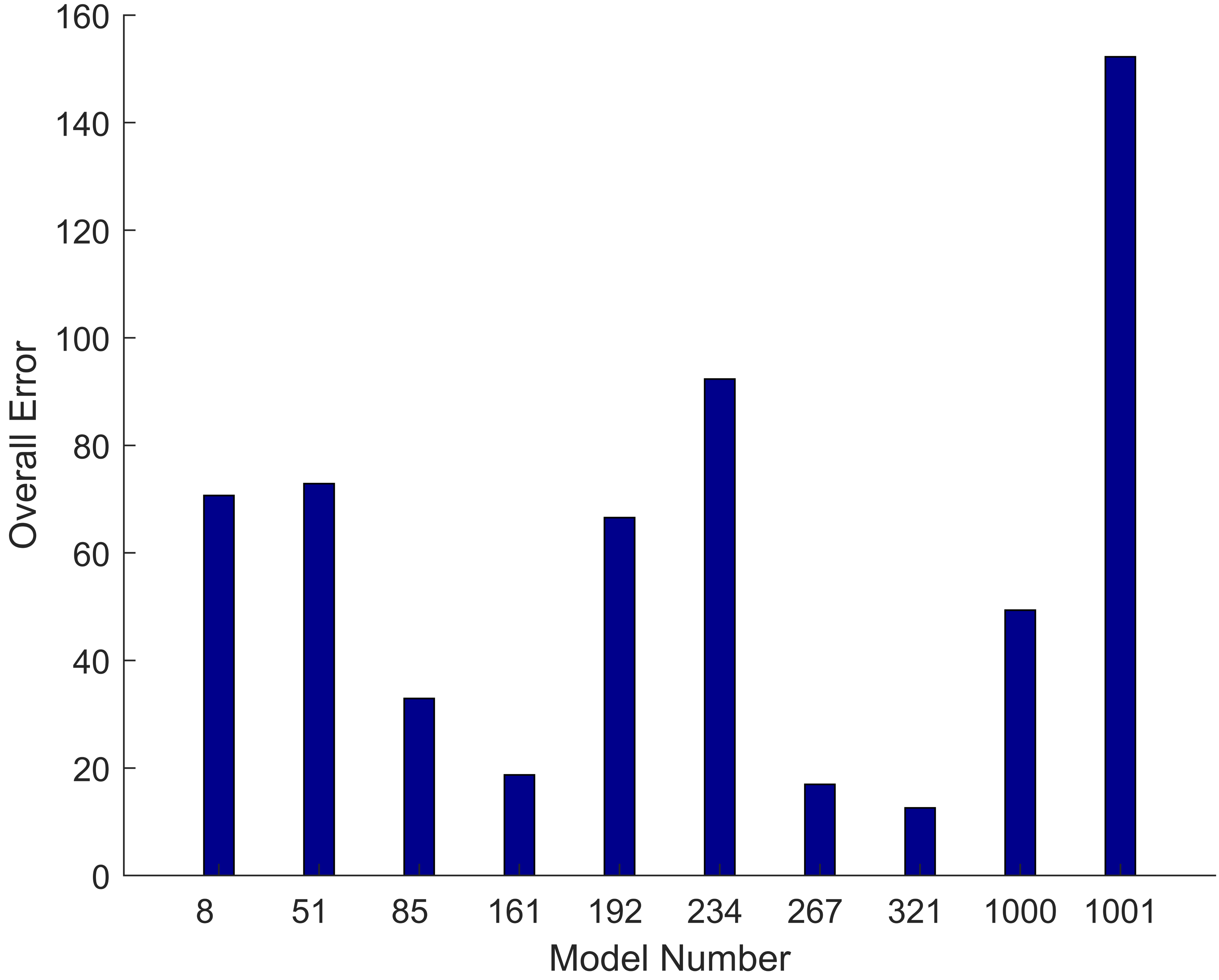}}
  \centerline{(c)}
\end{minipage}
\hfill
\begin{minipage}[b]{0.48\linewidth}
  \centering
  \centerline{\includegraphics[width=3.95cm]{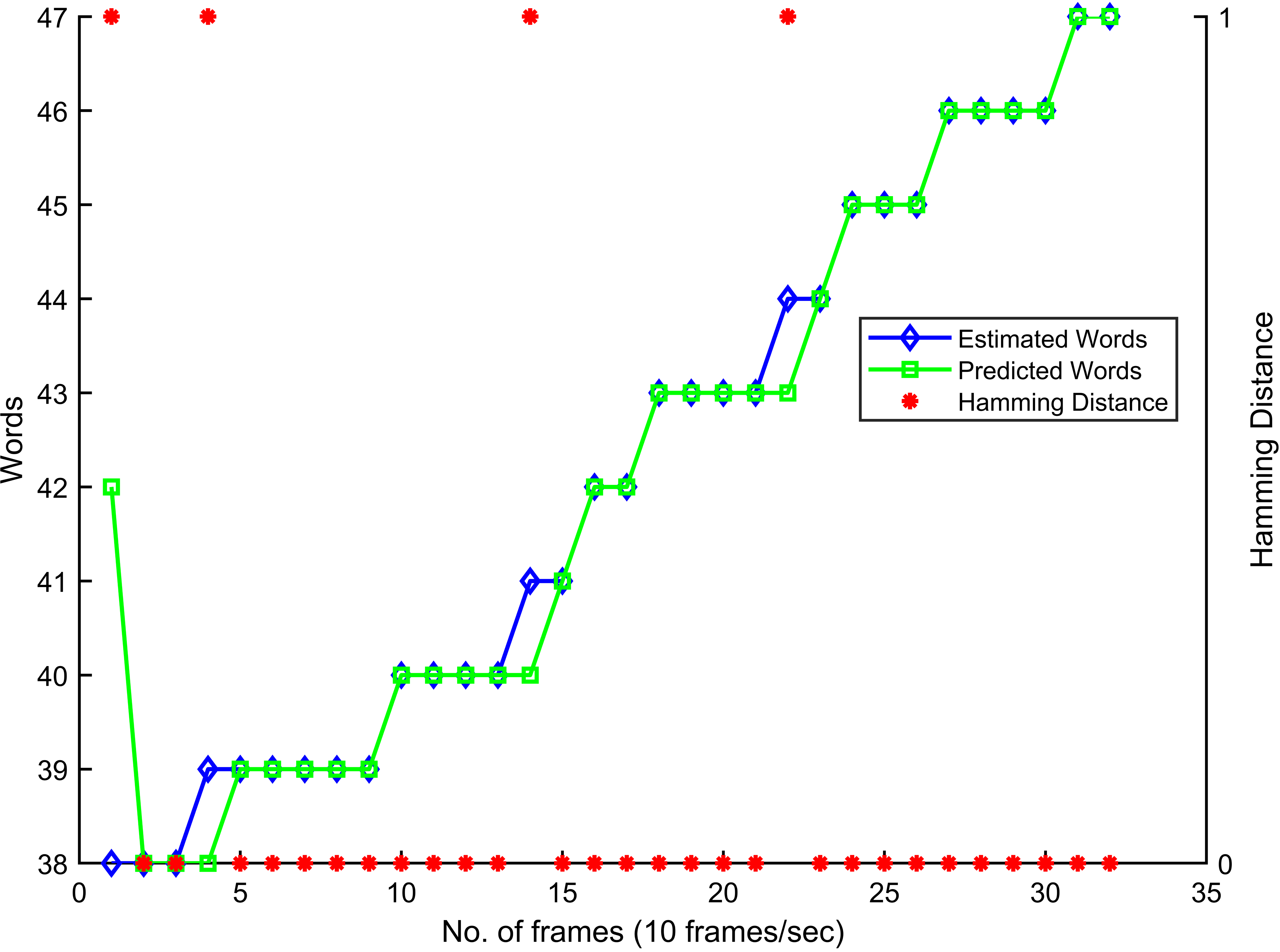}}
  \centerline{(d)}
\end{minipage}


\caption{Examples of word prediction under normal situation: (a) Selection of generative model \(51\), (b) Corresponding predictions, (c) Selection of generative models \(267\) \& \(321\), (d) Corresponding predictions.}
\label{fig.6}
\end{figure}

\begin{figure}[htb]
\begin{minipage}[b]{0.48\linewidth}
  \centering
  \centerline{\includegraphics[width=3.65cm]{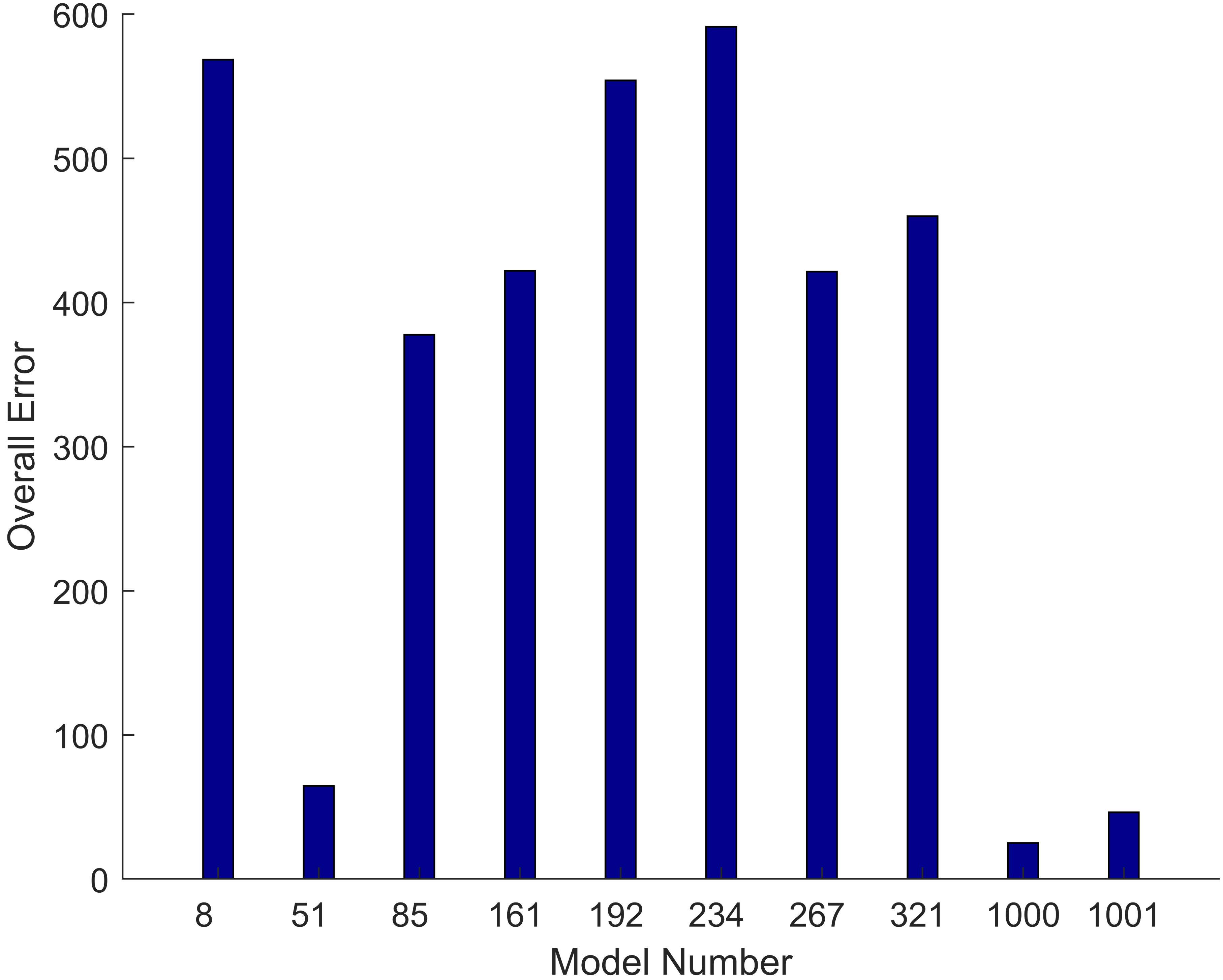}}
 \centerline{(a)}
\end{minipage}
\hfill
\begin{minipage}[b]{0.48\linewidth}
  \centering
  \centerline{\includegraphics[width=3.95cm]{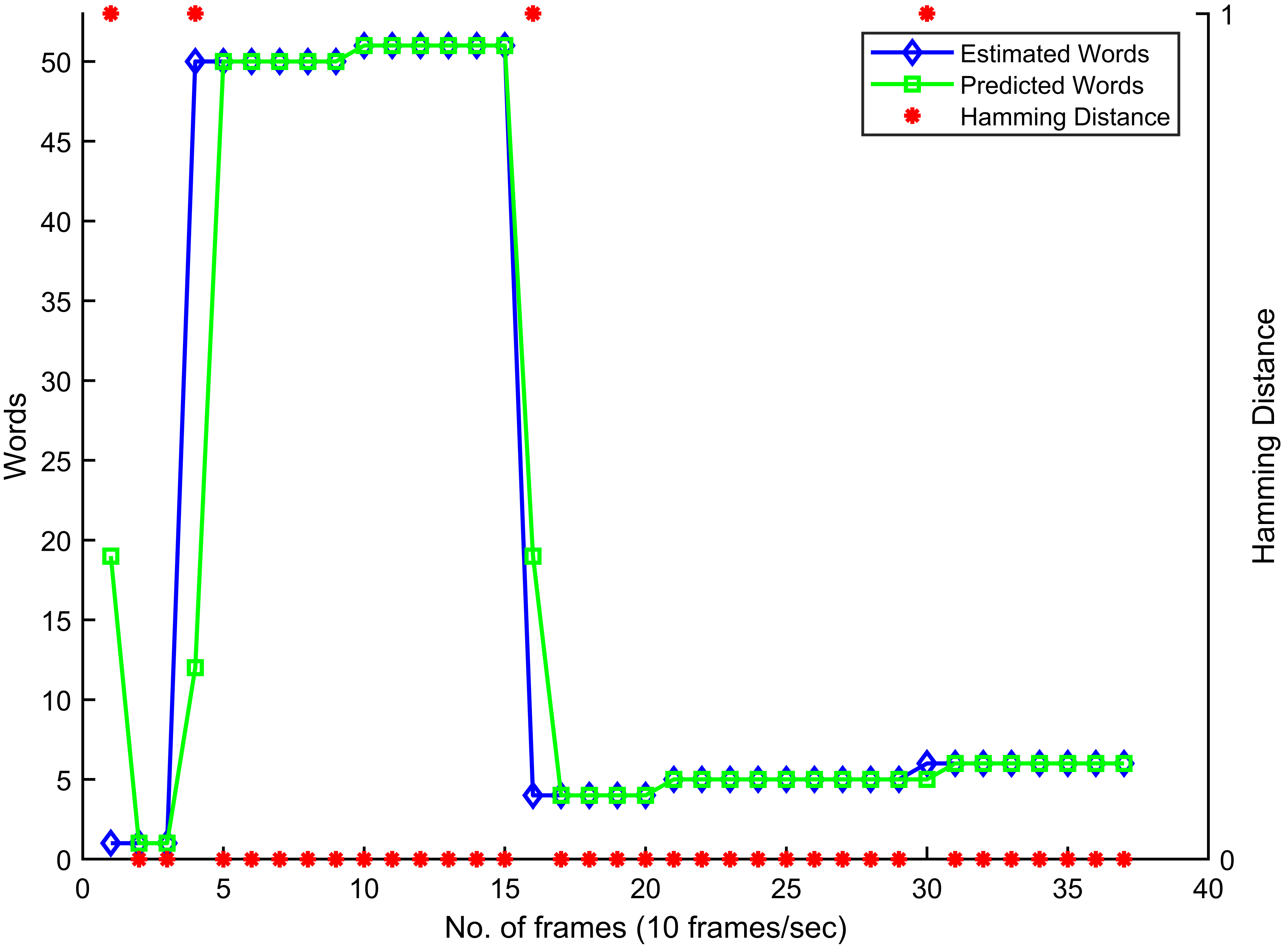}}
 \centerline{(b)}
\end{minipage}
\begin{minipage}[b]{0.48\linewidth}
  \centering
  \centerline{\includegraphics[width=3.65cm]{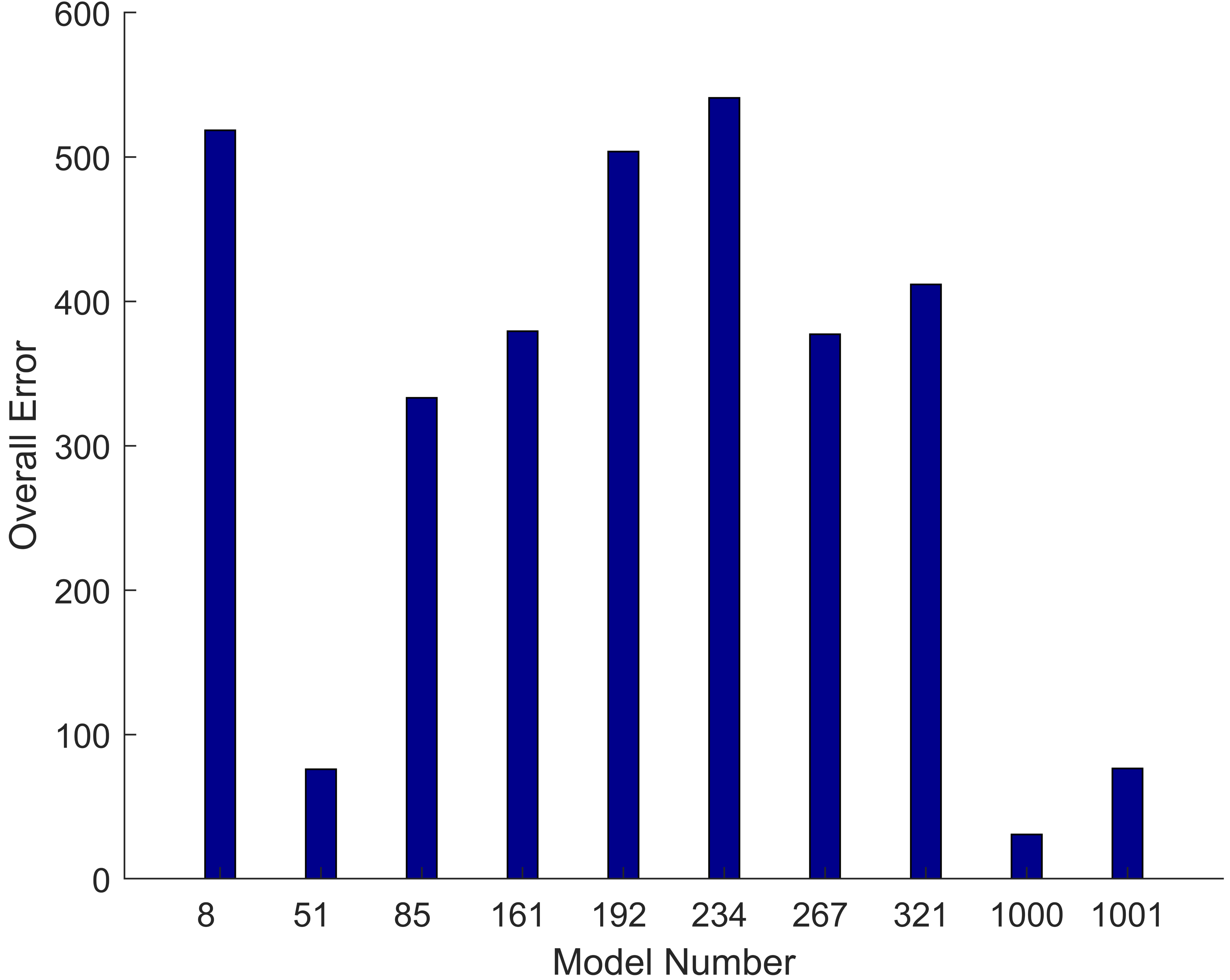}}
 \centerline{(c)}
\end{minipage}
\hfill
\begin{minipage}[b]{0.48\linewidth}
  \centering
  \centerline{\includegraphics[width=3.95cm]{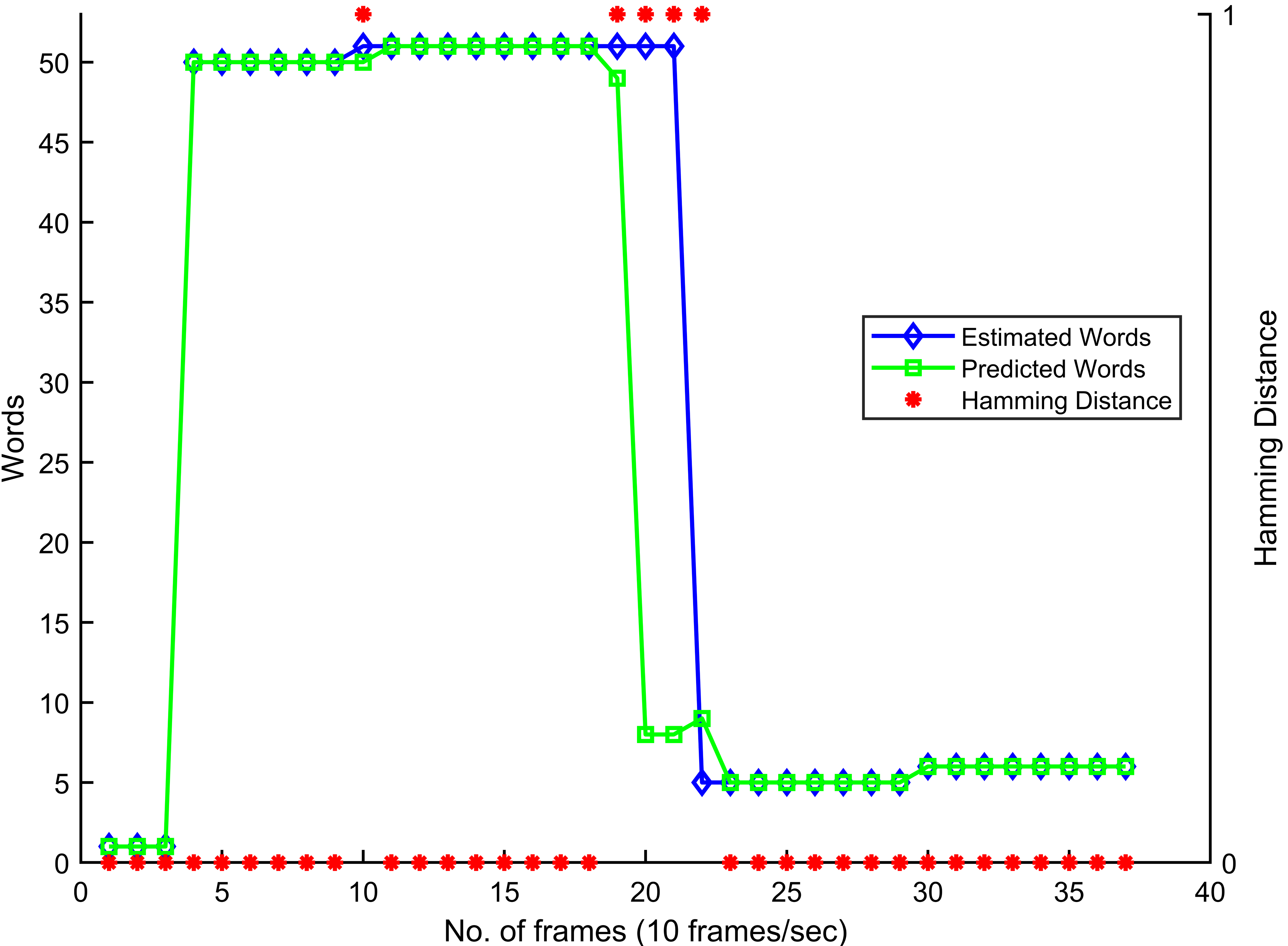}}
  \centerline{(d)}
\end{minipage}
\caption{Examples of word prediction under blockages situation: (a) Selection of generative model \(1000\), (b) Corresponding predictions, (c) Selection of generative model \(1000\) under the new situation, (d) Corresponding predictions.}
\label{fig.7}
\end{figure}

Fig.~\ref{fig.6} shows different examples of testing tracks under normal situations, with the selection of the best predictive model. We can observe that the predicted words match the estimated words with very less abnormalities. These abnormalities are caused due to errors in finding the estimated words (while finding the nearest clusters). Fig.~\ref{fig.6}(a) and Fig.~\ref{fig.6}(c) show the overall errors while finding the best generative models. The models with fewer errors, i.e., 51 for single a track and 267 and 321 for interactive tracks, are chosen for the predictions, respectively, which can also be seen in Fig.~\ref{fig.3}(b). 

Fig.~\ref{fig.7} shows the performance in the blockage and the new situation. We can observe that a blockage model ID-1000 is chosen in Fig.~\ref{fig.7}(a) for the blockage predictions which are shown in Fig.~\ref{fig.7}(b), this model gives less abnormalities because the I-GDBN was aware of the blockage at this location, and the model was already trained from the normal GDBN model ID-51. However, some abnormalities are detected in Fig.~\ref{fig.7}(d), when the duration of blockage exceeds, i.e., from frame 18 to 22, because the model was not aware of the blockage in this new situation. In this case, the selected blockage model is the same, i.e., GDBN model ID-1000 (see Fig.~\ref{fig.7}(c)).

\section{Conclusion}
\label{sec:conclusion&Futurework}
This paper proposes a new data-driven approach to learning an I-GDBN model from temporal 3D LiDAR point cloud perception for a self-aware agent. The proposed framework efficiently learns various normal and blockage models, performs interactions between multiple dynamic vehicles, and better predicts the LiDAR sensor blockages. Additionally, our work addresses optimal model selection, thus improving the prediction accuracy in various situations. Hence, the proposed approach is appropriate to support the self-aware and explainable characteristics of the proposed interactive model. 
This work can be extended by integrating additional data modalities, such as communication signals with 3D LiDAR for integrated sensing and communication (ISAC) systems. 

\section*{Acknowledgment}
This work was partially supported by the European Union under the Italian National Recovery and Resilience Plan (PNRR) of NextGenerationEU partnership on ``Telecommunications of the Future'' (PE$00000001$ - program ``RESTART''), CUP E6$3$C$22002040007$ - D.D. n.$1549$ of $11/10/2022$.



\end{document}